\documentclass[]{aastex631}

\usepackage{amsmath}

\def\apjl{ApJL}
\def\aap{A\&A}

\def\apjs{ApJ Supp}

\newcommand{\lsim }{{\lower0.8ex\hbox{$\buildrel <\over\sim$}}}
\newcommand{\gsim }{{\lower0.8ex\hbox{$\buildrel >\over\sim$}}}
\newcommand{\Msun}{\ifmmode {M_{\odot}}\else${M_{\odot}}$\fi}
\newcommand{\Lsun}{\ifmmode {L_{\odot}}\else${L_{\odot}}$\fi}
\newcommand{\Rsun}{\ifmmode {R_{\odot}}\else${R_{\odot}}$\fi}
\newcommand{\ergs}{erg~s$^{-1}$}

\newcommand{\coordra}[6]{R.A.(#1)$=#2^{h}~#3^{m}~#4\fs#5\pm#6$}
\newcommand{\coorddec}[6]{Dec.(#1)$=#2\degr~#3\arcmin~#4\farcs #5\pm#6$}

\newcommand{\xrt}{\textit{Swift}/XRT}

\newcommand{\chandra}{\textit{Chandra}}
\newcommand{\acis}{\textit{Chandra}/ACIS}

\newcommand{\maxi}{\textit{MAXI}}

\newcommand{\nustar}{\textit{NuSTAR}}

\newcommand\gc{GLIMPSE-C01}%
\newcommand\target{MAXI~J1848$-$015}%

\graphicspath{{./}{figures/}}

\newcommand{\distance}{3.4~kpc}
\newcommand{\distancewer}{$3.4\pm0.3$~kpc}

\newcommand{\midplacedistance}{6~pc}

\newcommand{\reddening}{4.85} 

\newcommand{\qLx}{$\leq 3.3\times10^{30}$~\ergs}

\newcommand{\vlitepaper}{Peters~et~al.,~\textit{in. prep.}}
\newcommand{\followuppaper}{Tremou~et~al.,~\textit{in. prep.}}

\begin{document}

\title{\target: The First Detection of Relativistically Moving Outflows from a Globular~Cluster X-ray Binary}

\correspondingauthor{Arash Bahramian}
\email{arash.bahramian@curtin.edu.au}

\author[0000-0003-2506-6041]{A. Bahramian}
\affil{International Centre for Radio Astronomy Research – Curtin University, GPO Box U1987, Perth, WA 6845, Australia}

\author[0000-0002-4039-6703]{E. Tremou}
\affil{National Radio Astronomy Observatory, Socorro, NM 87801 USA}

\author[0000-0003-3906-4354]{A. J. Tetarenko}
\affil{Department of Physics and Astronomy, Texas Tech University, Lubbock, TX 79409-1051, USA}
\affil{East Asian Observatory, 660 N. A`oh\={o}k\={u} Place, University Park, Hilo, Hawaii 96720, USA}
\affil{NASA Einstein Fellow}

\author[0000-0003-3124-2814]{J. C. A. Miller-Jones}
\affil{International Centre for Radio Astronomy Research – Curtin University, GPO Box U1987, Perth, WA 6845, Australia}

\author{R. P. Fender}
\affil{Astrophysics, Department of Physics, University of Oxford, Keble Road, Oxford OX1 3RH, UK}
\affil{Department of Astronomy, University of Cape Town, Private Bag X3, Rondebosch 7701, South Africa}

\author[0000-0001-5538-5831]{S. Corbel}
\affil{Université Paris Cité and Université Paris Saclay, CEA, CNRS, AIM, F-91190 Gif-sur-Yvette, France}
\affil{ORN, Observatoire de Paris, Université PSL, Université Orléans, CNRS, F-18330 Nançay, France}

\author[0000-0001-7361-0246]{D. R. A. Williams}
\affil{Jodrell Bank Centre for Astrophysics, School of Physics and Astronomy, The University of Manchester, Manchester, M13 9PL}

\author[0000-0002-1468-9668]{J. Strader}
\affil{Center for Data Intensive and Time Domain Astronomy, Department of Physics and Astronomy, Michigan State University, East Lansing, MI 48824, USA}

\author[0000-0002-0426-3276]{F. Carotenuto}
\affil{Astrophysics, Department of Physics, University of Oxford, Keble Road, Oxford OX1 3RH, UK}

\author[0000-0002-1206-1930]{R. Salinas}
\affil{Gemini Observatory/NSF's NOIRLab, Casilla 603, La Serena, Chile}

\author[0000-0002-6745-4790]{J. A. Kennea}
\affil{Department of Astronomy and Astrophysics, The Pennsylvania State University, 525 Davey Lab, University Park, PA 16802, USA}

\author[0000-0002-6154-5843]{S. E. Motta}
\affil{Istituto Nazionale di Astrofisica, Osservatorio Astronomico di Brera, via E. Bianchi 46, I-23807 Merate (LC), Italy}

\author[0000-0002-6896-1655]{P. A. Woudt}
\affil{Department of Astronomy, University of Cape Town, Private Bag X3, Rondebosch 7701, South Africa}

\author[0000-0002-3493-7737]{J. H. Matthews}
\affil{Astrophysics, Department of Physics, University of Oxford, Keble Road, Oxford OX1 3RH, UK}

\author[0000-0002-7930-2276]{T. D. Russell}
\affil{INAF, Istituto di Astrofisica Spaziale e Fisica Cosmica, Via U. La Malfa 153, I-90146 Palermo, Italy}



\begin{abstract}
Over the past decade, observations of relativistic outflows from outbursting X-ray binaries in the Galactic field have grown significantly. In this work, we present the first detection of moving and decelerating radio-emitting outflows from an X-ray binary in a globular cluster. \target\ is a recently discovered transient X-ray binary in the direction of the globular cluster \gc. Using observations from the VLA, and a monitoring campaign with the MeerKAT observatory for 500 days, we model the motion of the outflows. This represents some of the most intensive, long-term coverage of relativistically moving X-ray binary outflows to date. We use the proper motions of the outflows from \target\ to constrain the component of the intrinsic jet speed along the line of sight,  $\beta_\textrm{int} \cos \theta_\textrm{ejection}$, to be $=0.19\pm0.02$. Assuming it is located in \gc, at \distance, we determine the intrinsic jet speed, $\beta_\textrm{int}=0.79\pm0.07$, and the inclination angle to the line of sight, $\theta_\textrm{ejection}=76^\circ\pm2^{\circ}$. This makes the outflows from \target\ somewhat slower than those seen from many other known X-ray binaries. We also constrain the maximum distance to \target\ to be $4.3$ kpc. Lastly, we discuss the implications of our findings for the nature of the compact object in this system, finding that a black hole primary is a viable (but as-of-yet unconfirmed) explanation for the observed properties of \target. If future data and/or analysis provide more conclusive evidence that \target\ indeed hosts a black hole, it would be the first black hole X-ray binary in outburst identified in a Galactic globular cluster.
\end{abstract}

\keywords{Radio jets (1347), Low-mass X-ray binary stars (939), Neutron stars (1108), Black holes (162), Stellar accretion (1578), Globular star clusters (656)}


\section{Introduction}
On 2020-12-20, the Monitor of All-sky X-ray Image (\maxi) mission detected a bright outburst from the direction of the globular cluster \gc\ \citep{Takagi2020}. At the time, the source (named \target) was located in the Sun-constraint zone for most other X-ray observatories. However, follow-up observations by the Nuclear Spectroscopic Telescope Array (\nustar) improved on localization and found indications of spectral evolution \citep{Pike2020, Mihara2021}. A follow-up observation by the X-ray Telescope onboard the Neil Gehrels Swift Observatory (\xrt) on 2021-02-21 found \target\ still in outburst and localized it to the core of \gc, suggesting that this outburst may come from a previously undetected X-ray binary (XRB) in this star cluster \citep{Kennea2021}. A further follow-up observation by the \chandra\ observatory provided improved constraints on the location \citep{Chakrabarty2021}, while inspection of archival \chandra\ data provided a deep upper-limit on the quiescent X-ray luminosity of \qLx\ in the 0.5--10 keV band \citep{Hare2021}.

In contrast with the Galactic field, globular clusters are known to contain an overabundance of X-ray binaries \citep[XRBs][]{Clark1975}. This is generally attributed to dynamical formation of these systems through encounters as opposed to canonical binary evolution \citep{Sutantyo1975, Fabian1975, Hills1976}. Observational evidence in support of this overabundance has been discussed both for Galactic \citep{Pooley2003, Heinke2003c, Bahramian2013}, and extragalactic \citep{Sarazin2003, Kundu2007, Kundu2007a} clusters.

\gc\ is a star cluster located $0.1^{\circ}$ away from the Galactic plane \citep{Kobulnicky2005}. It is at an estimated distance of \distance\ \citep{Hare2018} from us, corresponding to a distance of \midplacedistance\ from the Galactic midplane. This highly extinguished cluster \citep[$E(B-V)\approx\reddening$,][version 2010]{Harris1996}, has been suspected to be a globular cluster (possibly passing through the Galactic plane or corotating with the Galactic disk; Salinas et al., {\it in prep.}) due to properties including its well-populated giant branch - indicating an abundance of old low-mass giants, as opposed to young supergiants, and high stellar density in the core -- hinting at mass segregation, a well-known property of old, relaxed globular clusters \citep{Kobulnicky2005, Ivanov2005, Davidge2016}. In the X-rays, \gc\ is known to host more than a dozen X-ray sources \citep{Pooley2007, Hare2018}.

In this work, we report the detection of radio outflows moving away from the core of \target, observed by the Karl G. Jansky Very Large Array (VLA) and MeerKAT. We also detect clear deceleration of these outflows as they move through the interstellar medium (ISM). Moving outflows have been observed in a number of black hole XRBs \citep{Mirabel1994, Hjellming1995, Fomalont2001a, Corbel2002, Corbel2005, Yang2011, Marti2017, Rushton2017, MillerJones2019, Bright2020}, and with the emergence of facilities such as MeerKAT, decelerating outflows interacting with the ISM have now been seen in several XRBs at different distances, opening angles and times (up to months or years). However, \target\ is the first XRB in a globular cluster to be observed exhibiting such outflows. Furthermore, the outflows from \target\ show some of the most persistent radio emission from relativistically-moving XRB outflows observed to date \citep[see also XTE~J1550$-$564, XTE~J1748$-$288, MAXI~J1535$-$571, MAXI~J1820+070, MAXI~J1348$-$630;][]{Corbel2001, Brocksopp2007, Russell2019, Bright2020, Carotenuto2021}. In \S\ref{sec:data} we describe the data presented in this work and details of the data processing. In \S\ref{sec:result}, we determine the  location of a potential compact radio core and measure the motion of the moving outflows, and finally in \S\ref{sec:disc} we explore the implications of our findings, considering the host environment and making comparisons with other, similar, systems. The outflows were also independently identified in data from the VLA Low-band Ionosphere and Transient Experiment (VLITE) at 340 MHz (\vlitepaper).

\section{Data and Reduction}\label{sec:data}

Soon after the localization of \target\ to the core of \gc, we began an intense monitoring campaign in the radio band to constrain its nature and study the outflows. The observations analyzed in this work are summarized in Table~\ref{tab:obs_list}. 

\subsection{VLA}\label{sec:vla}
\target\ was observed with the VLA (Project Code: 21A--400) on 2021-02-27 (MJD 59272), for a total on-source observation time of $\sim 88$ min. During our observations, the array was in its most extended A configuration, observing in the 8 -- 12 GHz band, with the 3-bit samplers. The correlator was set up to generate 2 base-bands, each with 16 spectral windows comprised of 64 2-MHz channels, giving a total bandwidth of 2.048 GHz per base-band.  We reduced and imaged the data within the Common Astronomy Software Application package, \textsc{casa} \citep[version 5.4,][]{CASA2022}, using standard procedures outlined in the \textsc{casa} Guides\footnote{\url{https://casaguides.nrao.edu/index.php/Karl\_G.\_Jansky_VLA_Tutorials}.} for VLA data reduction (i.e., a priori flagging, setting the flux density scale, initial phase calibration, solving for antenna-based delays, bandpass calibration, gain calibration, scaling the amplitude gains, and final target flagging). We used 3C286 (J1331+3030) as a flux/bandpass calibrator, and J1851+0035 as a phase calibrator. We flagged short-spacings ($<400{\rm k}\lambda$) for J1851+0035, as the VLA calibrator catalogue\footnote{\url{https://science.nrao.edu/facilities/vla/observing/callist\#section-0}} indicates it is extended at these short-spacings.
To image the target source, we used Briggs weighting with robust parameter 0 (to balance sensitivity and resolution), two Taylor terms to account for the wide fractional bandwidth, the multi-scale algorithm (deconvolving with Gaussians of FWHM 5 and 15 pixels, as well as point sources), and placed an outlier field on a bright source to the south of our target\footnote{Likely, GPS5~031.243$-$0.110, a candidate ultracompact HII region \citep{Becker1994}.}. A zoomed-in image of the target is displayed in Figure~\ref{fig:vlaimg} ({\it left} panel), where we clearly detect two bright radio lobe structures and potentially see indications of a weak (at $3.5\sigma$ significance) core component midway between the lobes (see \S\ref{sec:localization}).

\subsection{MeerKAT}\label{sec:meerkat}
The field of \target ~was first observed with MeerKAT on 2021-02-28 (MJD 59273) as part of the ThunderKAT \footnote{\url{http://www.thunderkat.uct.ac.za/}} \citep[\textbf{T}he \textbf{HUN}t for \textbf{D}ynamic and \textbf{E}xplosive \textbf{R}adio transients with Meer\textbf{KAT},][]{2017fender} large survey project. Following the radio detection of \target ~\citep{Tremou2018}, we observed the field weekly for 15 minutes of on-source integration time. Due to the slow evolution of the outflows, in October 2021 we decreased the observations and reduced the monitoring cadence to bi-weekly (for a total of 54 observations). Our observations were made using the L-band receiver, centered at 1.284 GHz with a bandwidth of 0.856 GHz.  Observations typically alternated between the target and phase calibrator (J1911$-$2006), with a single scan on a bandpass and flux calibrator, J1939$-$6342. All observations were obtained in full polarization mode. We flagged the data using \textsc{Tricolour} \citep{2022hugo} and calibrated the data using \textsc{casa} as part of the \textsc{Oxkat} data reduction scripts \citep{2020oxkat}. For imaging the field of \target, we used \textsc{WSClean}\citep{2014offringa}. The image size was 10240 $\times$ 10240 pixels, the pixel size was set at 1.1$\arcsec$ and we used Briggs weighting with a robust parameter of $-0.6$. We used the \textsc{casa} task \texttt{imfit} to extract the flux densities and the positions of the two lobes. The two components were separated significantly and a Gaussian fit (constrained to the size of the synthesized beam) was applied at each ejection component. We also account for a 10\%  of the synthesized beam as a positional error.

\begin{longtable}[c]{ccc|ccc|ccc} 
\caption{VLA and MeerKAT observations of \target\ discussed in this work. MJD column represents the modified Julian date at the start of the observation. The VLA X-band observation (performed on 2021-02-27) was centered at 10 GHz. The MeerKAT L-band observations were all centered at 1.284 GHz.}
\label{tab:obs_list}
\endfirsthead                                                  
\hline             
Date        & MJD      & Exposure & Date        & MJD      & Exposure & Date        & MJD      & Exposure \\
\hline
\hline
2021-02-27 & 59272.522 & 4 hr   & 2021-07-04 & 59399.901 & 15 min & 2021-12-18 & 59566.443 & 15 min \\
2021-03-06 & 59279.221 & 15 min & 2021-07-12 & 59407.019 & 15 min & 2022-01-03 & 59582.381 & 15 min \\
2021-03-13 & 59286.110 & 15 min & 2021-07-26 & 59421.702 & 15 min & 2022-01-16 & 59595.315 & 15 min \\
2021-03-20 & 59293.150 & 15 min & 2021-07-31 & 59426.898 & 15 min & 2022-01-29 & 59608.433 & 15 min \\
2021-03-28 & 59301.290 & 15 min & 2021-08-07 & 59433.909 & 15 min & 2022-02-14 & 59624.221 & 15 min \\
2021-04-05 & 59309.185 & 15 min & 2021-08-15 & 59441.921 & 15 min & 2022-02-27 & 59637.360 & 15 min \\
2021-04-10 & 59314.093 & 15 min & 2021-08-22 & 59448.673 & 15 min & 2022-03-16 & 59654.239 & 15 min \\
2021-04-19 & 59323.003 & 15 min & 2021-08-28 & 59454.676 & 15 min & 2022-03-28 & 59666.106 & 15 min \\
2021-04-24 & 59328.024 & 15 min & 2021-09-05 & 59462.751 & 15 min & 2022-04-10 & 59679.088 & 15 min \\
2021-05-01 & 59335.077 & 15 min & 2021-09-13 & 59470.646 & 15 min & 2022-04-25 & 59694.105 & 15 min \\
2021-05-07 & 59341.109 & 15 min & 2021-09-20 & 59477.678 & 15 min & 2022-05-07 & 59706.052 & 15 min \\
2021-05-15 & 59349.920 & 15 min & 2021-09-27 & 59484.594 & 15 min & 2022-05-23 & 59722.951 & 15 min \\
2021-05-22 & 59356.032 & 15 min & 2021-10-04 & 59491.658 & 15 min & 2022-06-03 & 59733.961 & 15 min \\
2021-05-27 & 59361.965 & 15 min & 2021-10-17 & 59504.508 & 15 min & 2022-06-17 & 59747.929 & 15 min \\
2021-06-05 & 59370.907 & 15 min & 2021-10-23 & 59510.519 & 15 min & 2022-07-03 & 59763.962 & 15 min \\
2021-06-12 & 59377.956 & 15 min & 2021-11-08 & 59526.638 & 15 min & 2022-07-15 & 59775.849 & 15 min \\
2021-06-19 & 59384.910 & 15 min & 2021-11-19 & 59537.677 & 15 min & 2022-07-30 & 59790.750 & 15 min \\
2021-06-27 & 59392.034 & 15 min & 2021-12-04 & 59552.425 & 15 min & \\
\hline
\end{longtable}

\section{Analysis and Results}\label{sec:result}
\subsection{Initial identification of moving lobes and a possible compact core}\label{sec:localization}

\begin{figure}
\centering
\includegraphics[width=3.5in]{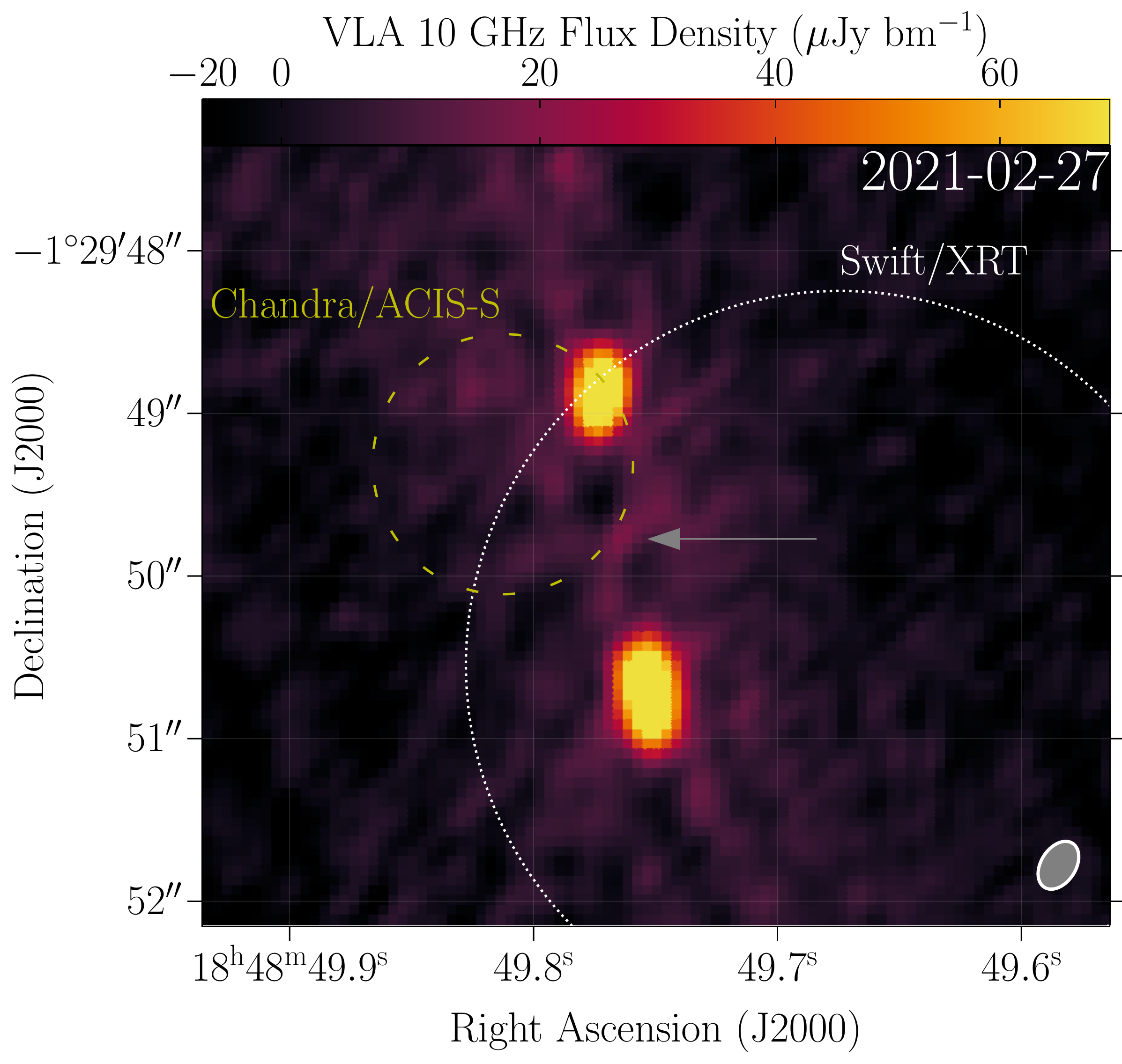}
\includegraphics[width=3.5in]{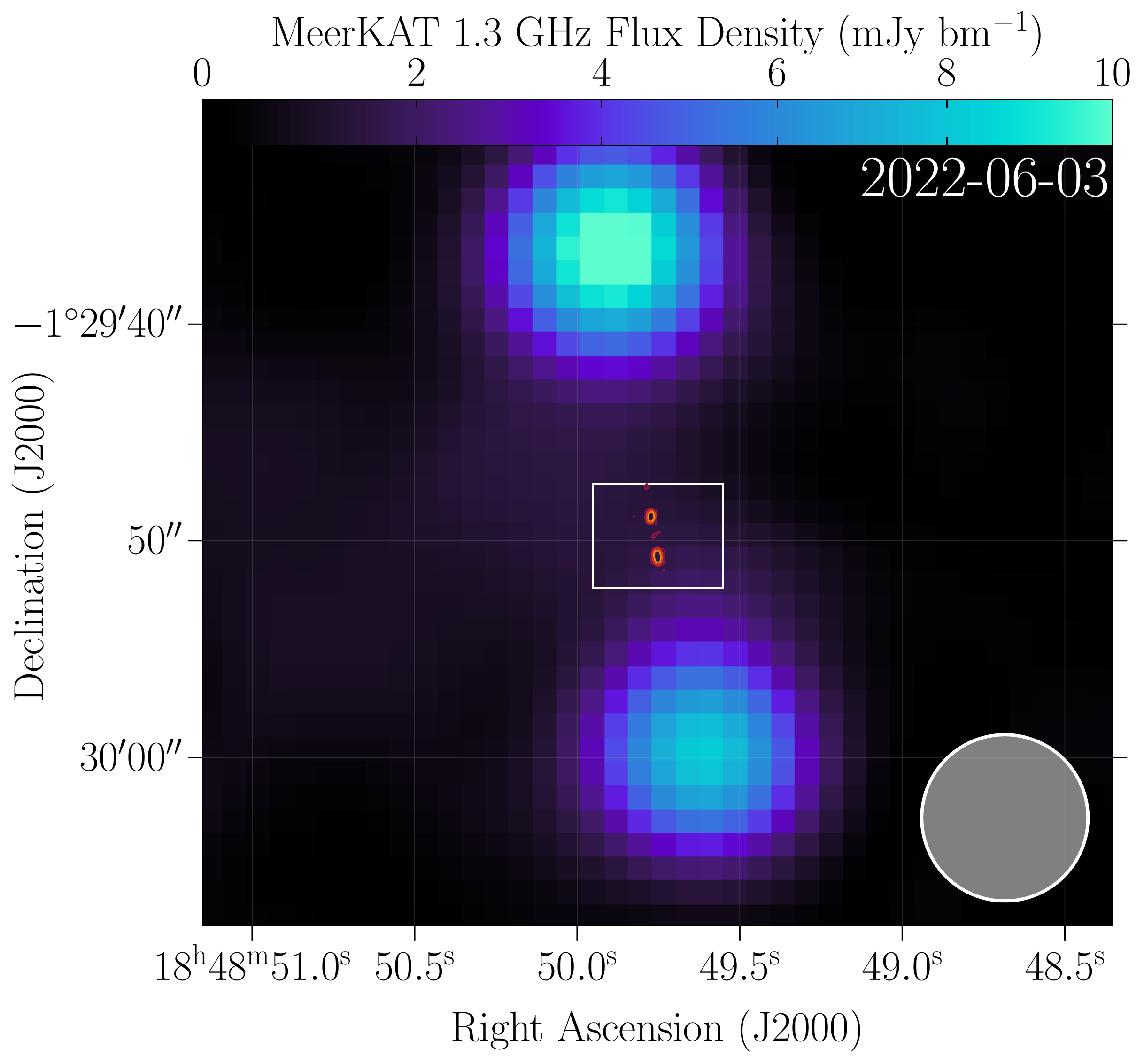}
\caption{\textit{Left}: VLA 10 GHz image of \target\ as observed on 2021-02-27. Two bright lobes are clearly detected. At the midpoint between the two lobes, a low-significance source (at $3.5\sigma$) is present (denoted by a gray arrow). We demonstrate the motion of these lobes in \S\ref{sec:motion}. The white dashed circle indicates the localization of the X-ray transient by \citet{Kennea2021} based on \xrt, while the yellow dotted circle represents the localization by \citet{Chakrabarty2021} based on \acis\ (both plotted with 90\% confidence localization radii). The VLA synthesized beam is represented with a filled gray ellipse in the corner of the image. \textit{Right}: MeerKAT 1.3 GHz image of outflows from \target\ as observed on 2022-06-03 (as an example for comparison with the VLA image on the left). The MeerKAT synthesized beam is represented with a filled gray ellipse in the corner of the image. The white box in the center indicates the region shown in the left panel and the red/orange contours are representative of the VLA flux densities on 2021-02-27 as shown in the left panel (representing 15, 30 and 60 $\mu$Jy, from red to yellow). Note that the two panels have different spatial and flux density scales.}
\label{fig:vlaimg}
\end{figure}

Our first radio observation of \target, performed by the VLA on 2021-02-27, revealed two bright sources with flux densities of $256\pm11$ (``Northern lobe'') and $176\pm6$ $\mu$Jy (``Southern lobe'') near the reported position of the system in the X-rays (Figure~\ref{fig:vlaimg} - {\it left}), with coordinates:

\begin{center}
Northern lobe:\quad\coordra{ICRS}{18}{48}{49}{7729}{0.002},\quad\coorddec{ICRS}{-01}{29}{48}{886}{0.03}

Southern lobe:\quad\coordra{ICRS}{18}{48}{49}{7532}{0.002},\quad\coorddec{ICRS}{-01}{29}{50}{725}{0.03}
\end{center}

As we discuss below in \S\ref{sec:motion}, our follow-up MeerKAT monitoring campaign allowed us to establish that these two are outward-moving lobes. Both of these sources appear partially resolved when compared to the synthesized beam, indicating expansion of the outflows. 

In addition to the bright lobes, we also notice a faint source approximately half-way between the two lobes, with a flux density of 17.6 $\mu$Jy, corresponding to a significance of $3.5\sigma$, when compared to the local noise R.M.S of 5.0 $\mu$Jy (calculated from an annulus around the source). Coordinates of this faint source (as estimated from the VLA observation on 2021-02-27) are:

\begin{center}
Potential compact core:\quad\coordra{ICRS}{18}{48}{49}{7628}{0.002},\quad\coorddec{ICRS}{-01}{29}{49}{758}{0.03}
\end{center}

The location of this faint source is consistent with reported localizations of \target\ in the X-rays by \acis\ and \xrt, indicating that it is likely the compact core of \target. Thus, hereafter, we assume this to be the case.

\subsection{Modeling the motion of the outflows}\label{sec:motion}

\begin{figure}
\centering
\includegraphics[width=6.5in]{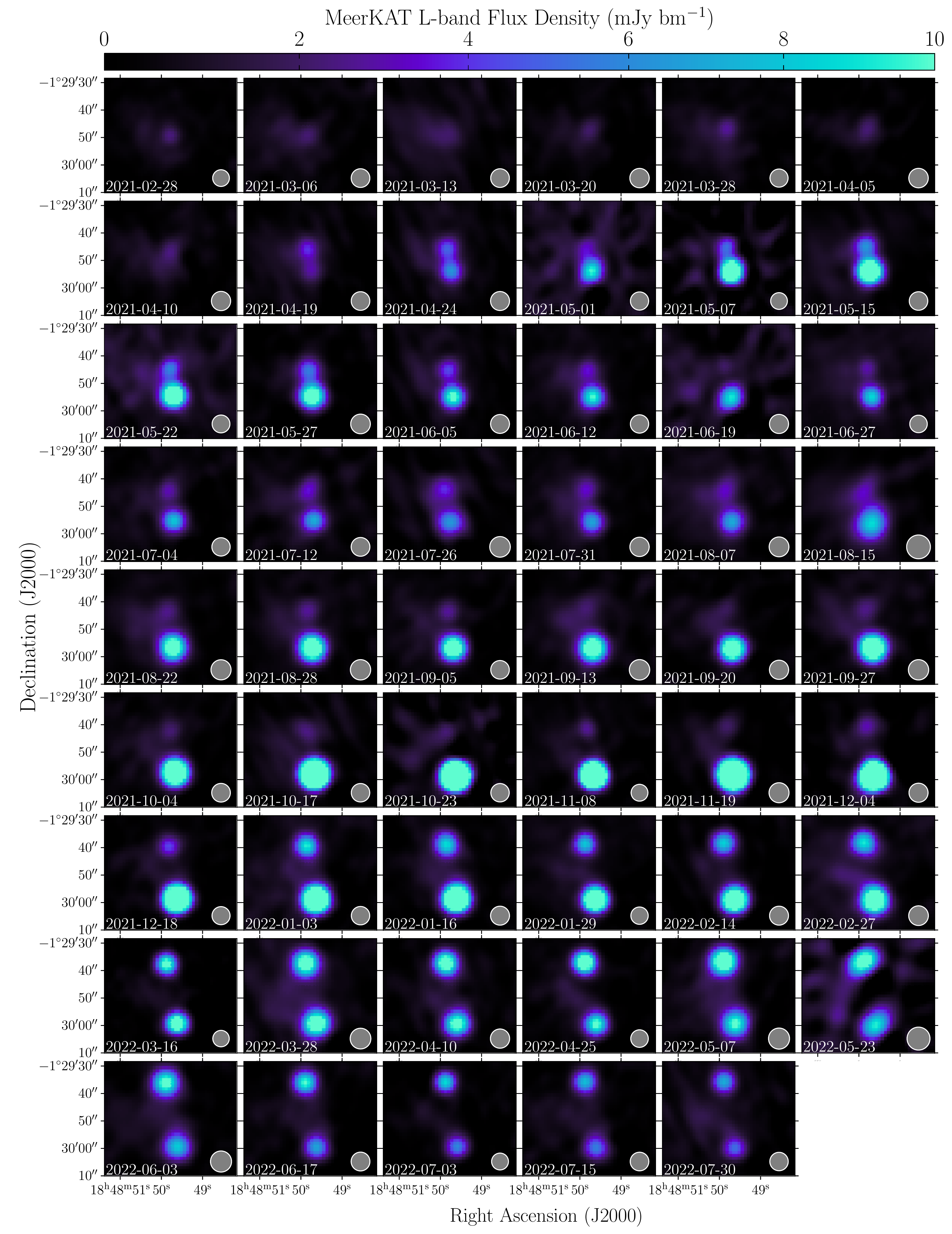}
\caption{Detection and evolution of the outflows from \target\ as seen by MeerKAT (in L-band at 1.28 GHz), since it was first detected in February 2021. Following the initial detection, both south and north lobes showed varying brightness. The two lobes were clearly distinguishable for the first time in April 2021, approximately two months after the initial detection of the lobes by the VLA.}
\label{fig:meerkatimg}
\end{figure}

The VLA in its most extended A-configuration provides a resolution (the full-width at half-maximum of the synthesized beam width) of $\sim0\farcs2$ at 10 GHz, whereas MeerKAT at 1.28~GHz provides a resolution of $\sim6''$. Thus, while the lobes were already distinguishable in the VLA observation in February 2021, it was not until April 2021 that the lobes were resolved by MeerKAT (Figure~\ref{fig:meerkatimg}). Prior to this, MeerKAT only detected a single source with a flux density of $\approx2$ mJy at a location consistent with the X-ray and VLA positions of \target. It is likely that this source is confused, as in addition to MeerKAT's inability to resolve the two sources at early times, contributions from pulsars, which can be abundant in globular clusters, may provide additional confusing sources at the frequency and resolution of our MeerKAT observations. The subsequent brightening of the lobes as they rose to a peak of $\sim 50$~mJy (Figure~\ref{fig:meerkatimg}), led to distinguishable detection of the two lobes by MeerKAT and enabled us to track the motion of the lobes and constrain their properties.  A full analysis of the flux density evolution of the lobes will be presented in \followuppaper.

We model the outflowing lobes with a kinematic model along their direction of motion, comprised of proper motion with constant deceleration, and eventual stalling at zero velocity. To do so, we first apply the following coordinate transformation to the location of the lobes in each observation:

\begin{equation}
\begin{bmatrix}
\theta^{N,i}_\perp\\
\theta^{N,i}_\parallel
\end{bmatrix}
=
\begin{bmatrix}
    \cos\phi & -\sin\phi\\
    \sin\phi & \cos\phi
\end{bmatrix}
\begin{bmatrix}
\alpha^{N,i}-\alpha^0\\
\delta^{N,i}-\delta^0\\
\end{bmatrix}
\end{equation}

where $\alpha$ and $\delta$ are right ascension and declination, with $[\alpha^{N,i}, \delta^{N,i}]$ representing coordinates of the northern lobe in each individual observation $i$. $[\alpha^0, \delta^0]$ represents the coordinates of the compact core, and $\phi$ is the jet axis on the sky plane - the angle between the lobes as initially measured in the first VLA observation ($9.1^{\circ}$ eastward from north in the equatorial coordinate system, \S\ref{sec:localization}). This transformation yields the displacement of the lobes as a combination of a component parallel to the jet axis ($\theta^{N,i}_\parallel$) and one perpendicular to that axis ($\theta^{N,i}_\perp$), allowing a simple ballistic modeling of the motion along the jet axis. An identical transformation is applied to the co-ordinates of the southern lobe in each observation. This transformation allows us to track the motions of the lobes along the jet axis while limiting the influence of any potential scatter as the lobes expand and their weighted mean position becomes dominated by any hot spots. Such a scatter is expected, given the resolved nature of the lobes in our initial VLA epoch. While they may remain unresolved by MeerKAT, their measured positions will be a weighted mean of the emission, which could be biased by a hot spot in an expanding lobe (\followuppaper).

\begin{figure}
\centering
\includegraphics[height=3.8in]{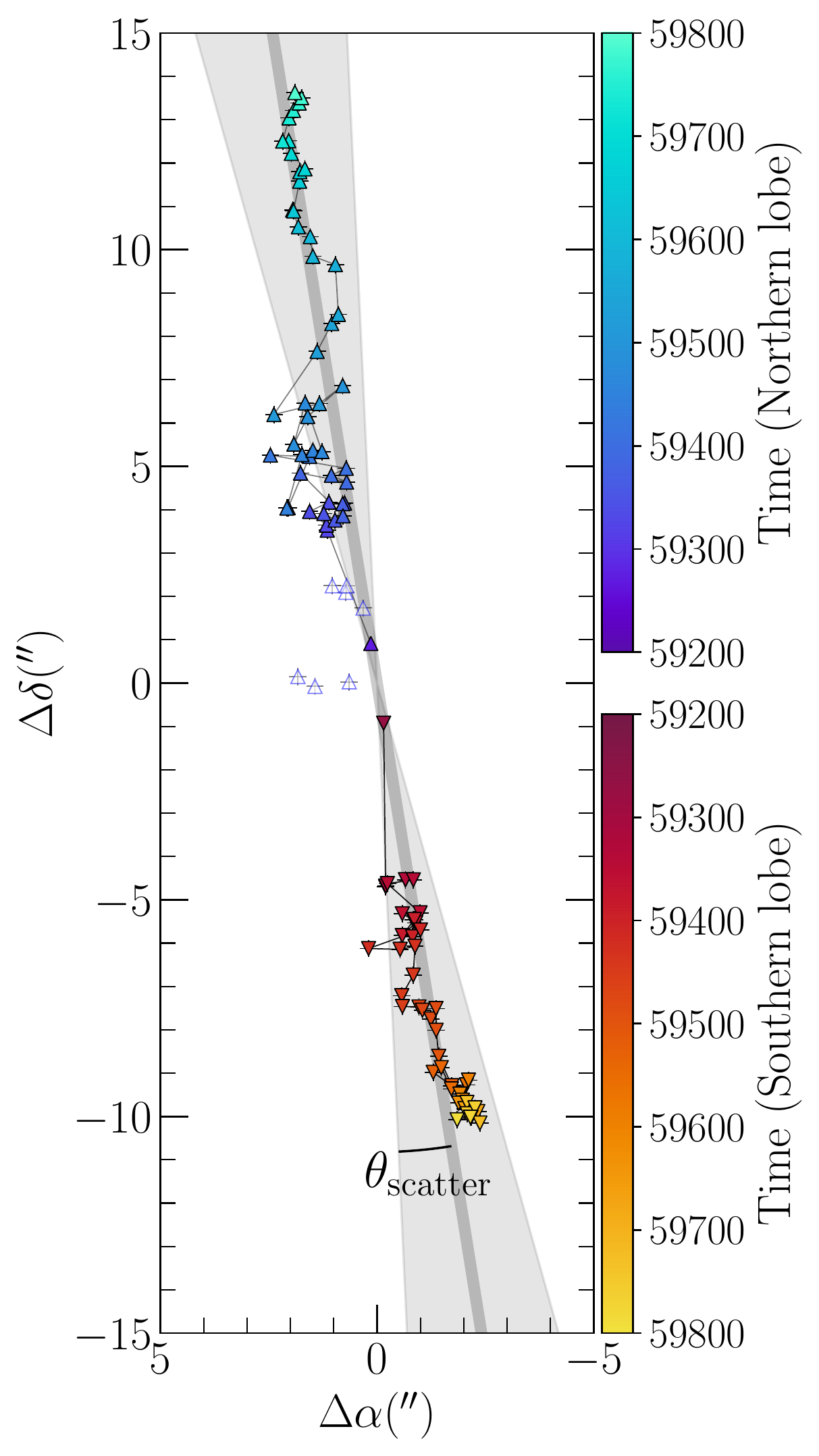}
\includegraphics[width=4.7in]{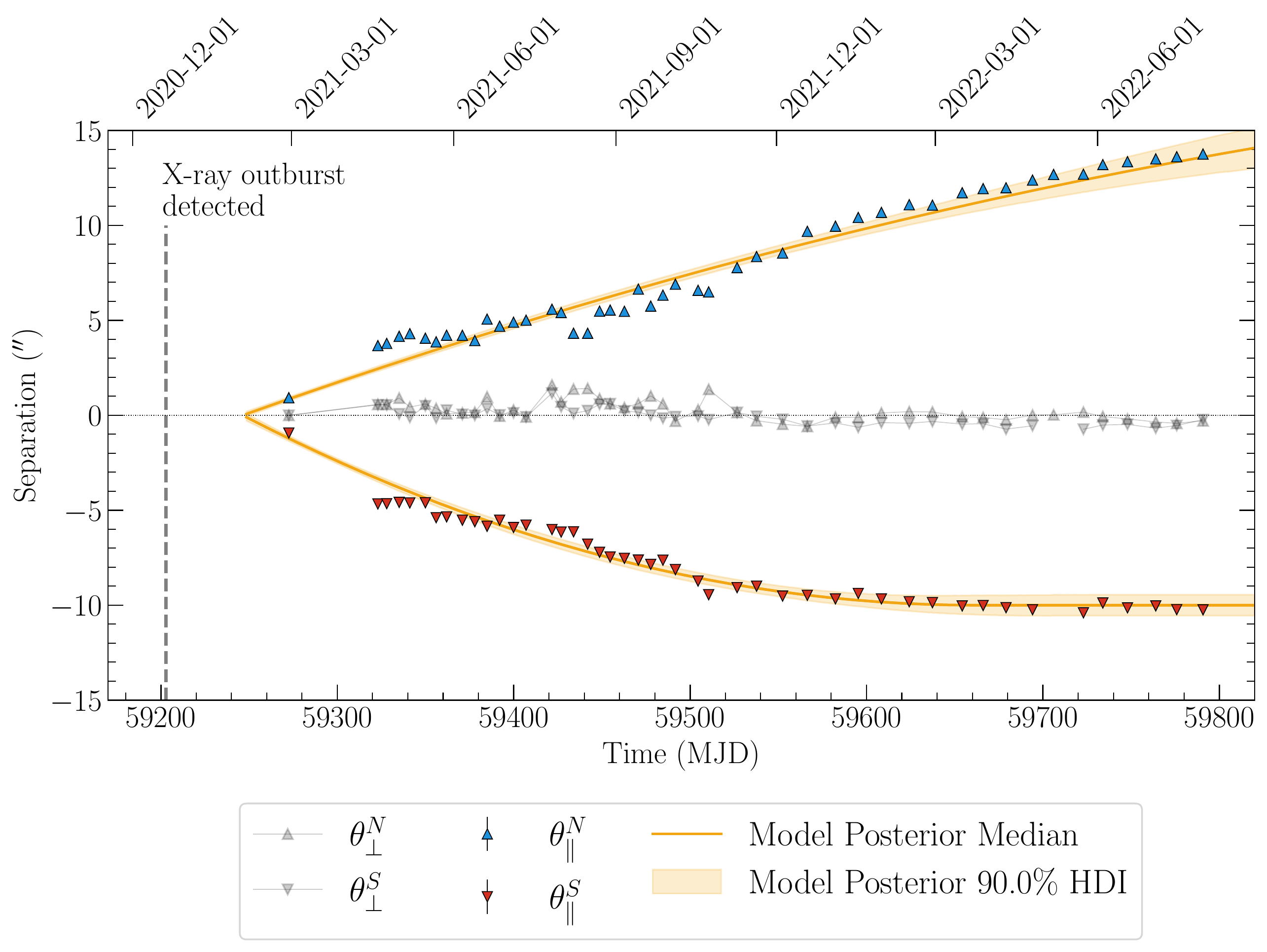}
\caption{{\it Left}: Position of the outflows from \target\ as observed over time.  Unfilled markers represent observations in which the lobes could not be distinguished clearly. These observations were not included in the modeling. The gray line represents the jet axis on the sky plane as determined by the location of the lobes in the first VLA observation. The shaded gray interval represents the scatter angle (posterior median) as determined by our model (\S\ref{sec:motion}). {\it Right}: Proper motion of the radio lobes parallel to the jet axis (colored triangles) and perpendicular to the jet axis (gray triangles), over time as observed during our monitoring campaign (errorbars are smaller than the markers). The yellow line indicates a best-fit model, based on posterior medians, and the shaded yellow interval represent the 90\% highest density interval of model posteriors.}
\label{fig:motion}
\end{figure}

\begin{figure}
\centering
\includegraphics[width=7in]{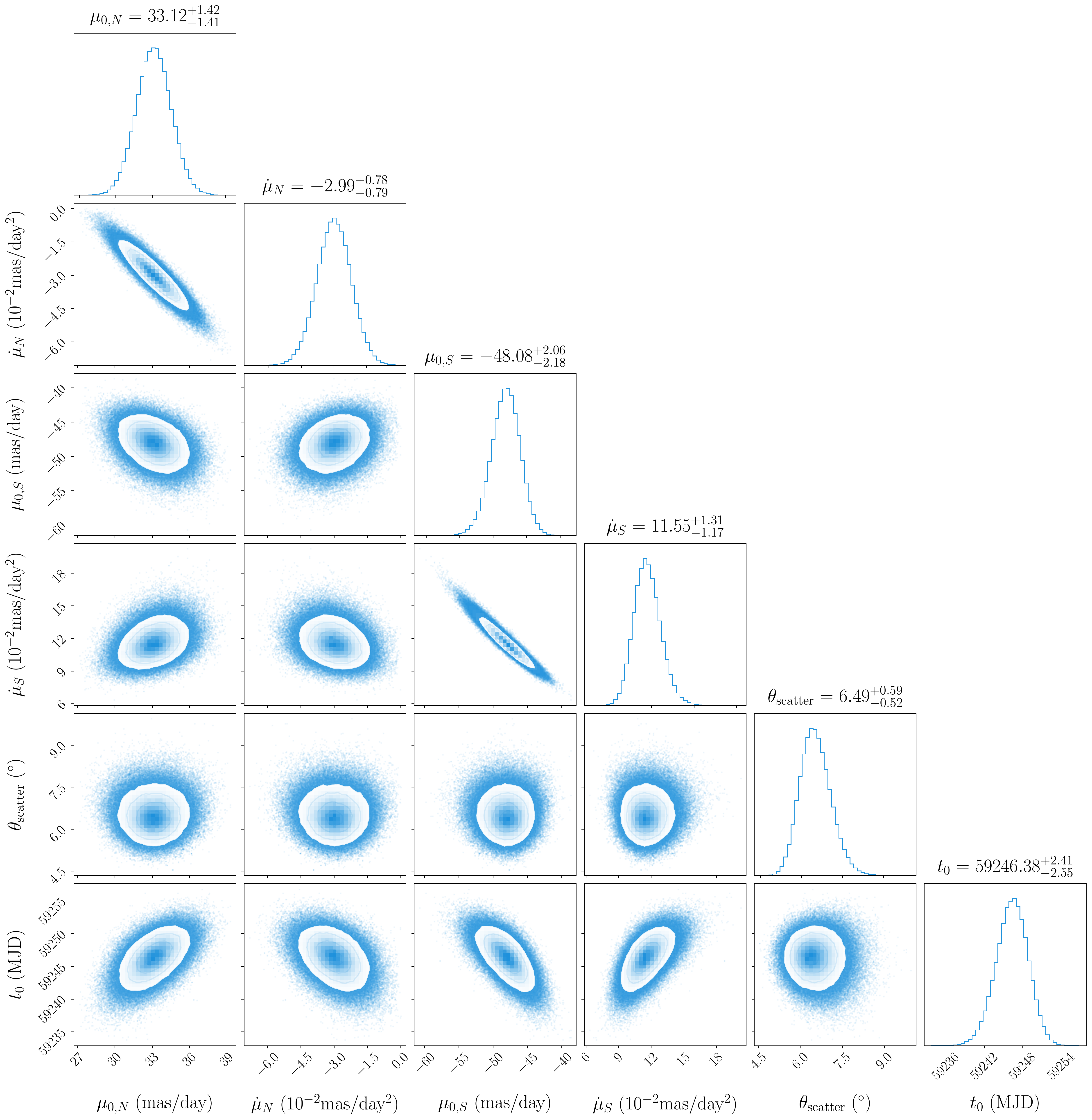}
\caption{Pairwise plots of the posterior samples for parameters in the model described in \S\ref{sec:motion}. All samples appear unimodal. Significant anti-correlations are noticeable between proper motions and decelerations of the lobes. This is expected as with a lower initial proper motion, less deceleration would be needed to achieve a similar displacement. Initial proper motions and decelerations also show some degree of (anti-)correlation with the predicted outflow launch date. This is likely due to the fact that the very early behavior of the outflows is only captured by a single highly influential observation (the VLA observation in February 2021, see Figure~\ref{fig:motion}).}
\label{fig:pairwise}
\end{figure}

With the transformed coordinates, we introduce the following model:
\begin{align}
    \theta_\parallel^N(t) &= 
        \begin{cases}
        \mu_{0,N} (t-t_0) + \frac{\dot{\mu}_N}{2}(t-t_0)^2\quad &t < t_{N,\textrm{stop}}\\
        \mu_{0,N} (t_{N,\textrm{stop}}-t_0) + \frac{\dot{\mu}_N}{2}(t_{N,\textrm{stop}}-t_0)^2\quad &t \geq t_{N,\textrm{stop}} \\
    \end{cases}\\
    \theta_\parallel^S(t) &= 
        \begin{cases}
        \mu_{0,S} (t-t_0) + \frac{\dot{\mu}_S}{2}(t-t_0)^2\quad &t < t_{S,\textrm{stop}}\\
        \mu_{0,S} (t_{S,\textrm{stop}}-t_0) + \frac{\dot{\mu}_S}{2}(t_{S,\textrm{stop}}-t_0)^2\quad &t \geq t_{S,\textrm{stop}}, \\
    \end{cases}
\end{align}
where $\mu_{0,S}$ and $\mu_{0,N}$ are proper motions at launch, $t_0$ is launch time, $\dot{\mu}_S$ and $\dot{\mu}_N$ are decelerations, and $t_{N,\textrm{stop}} = t_0 - \mu_{0,N} / \dot{\mu}_N$ and $t_{S,\textrm{stop}} = t_0 - \mu_{0,S} / \dot{\mu}_S$ are the times when the lobes stall completely due to deceleration. Given that the localizations of the northern and southern lobes are determined independently and that the lobes potentially move through different environments, we implement this model of motion with the  normal likelihood
\begin{align}
p(\theta^{N,i}_\parallel,\theta^{S,i}_\parallel|\mu_{0,N},\dot{\mu}_N,\mu_{0,S},\dot{\mu}_S,\theta_\textrm{scatter},t_0) = 
\mathcal{N}(\theta^{N,i}_\parallel|\theta_\parallel^N(t_i), \sigma^{N,i})
\mathcal{N}(\theta^{S,i}_\parallel|\theta_\parallel^S(t_i), \sigma^{S,i}),
\end{align}
where $\mathcal{N}$ denotes a normal probability density function,  $t_i$ is the time of the observation $i$, and
\begin{align}
\sigma^{N,i} = \sqrt{ (\sigma^{N,i}_{\theta,\parallel})^2 + (\theta^{N,i}_\parallel \tan \theta_\textrm{scatter})^2 }\\
\sigma^{S,i} = \sqrt{ (\sigma^{S,i}_{\theta,\parallel})^2 + (\theta^{S,i}_\parallel \tan \theta_\textrm{scatter})^2 }.
\end{align}
Here, $\sigma^{N,i}_{\theta,\parallel}$, $\sigma^{S,i}_{\theta,\parallel}$ are the statistical uncertainties on localizations of the lobes in the $\theta_\parallel$ component. The second term considers an additional localization uncertainty caused by the expansion of the lobes, characterized by $\theta_\textrm{scatter}$, which is the scatter angle as measured from the core along the jet axis (Figure~\ref{fig:motion} - {\it left}). In our model we assume that the motion of the northern and southern lobes are independent of each other in most aspects except the time of ejection ($t_0$) and the scatter angle ($\theta_\textrm{scatter}$). Using the same scatter angle for both lobes implies that the expansion of the lobes is dictated by their initial conditions. We assumed uniform priors for all model parameters as tabulated in Table~\ref{tab:parameters}.\footnote{We also verified that other choices of uninformative or weakly informative priors (such as log-uniform priors on $\mu$ and $\dot{\mu}$, or broad gaussian priors for $\theta_\textrm{scatter}$) do not influence the results.} We performed inference using Hamiltonian Monte-Carlo with No-U-turn sampling through \textsc{pymc} \citep[version 4.1.2,][]{Hoffman2011, Salvatier2016}. We verified chain convergence through the rank-normalized convergence diagnostic \citep{Gelman1992, Vehtari2019}, finding $\widehat{R} < 1.0001$ for each model parameter. Model parameter posterior samples are visualized in Figure~\ref{fig:pairwise}, with point and interval estimations reported in Table~\ref{tab:parameters}. The posterior sample median for $\theta_\textrm{scatter}$ is represented in Figure~\ref{fig:motion} - {\it left}, and the median and 90\% highest density interval of the posterior models are plotted against the data in Figure~\ref{fig:motion} - {\it right}.

The model above allows us to estimate the initial proper motion of the lobes ($\mu_{0,N}=33.1\pm1.4$ mas~d$^{-1}$ and $\mu_{0,S}=48.1\pm2.1$ mas~d$^{-1}$), and provides a constraint on the possible outflow launch time ($59246.4_{-2.5}^{+2.4}$ MJD, corresponding to 2021-02-01). The inferred launch date suggests the launch of outflows may have occurred $43.1_{-2.5}^{+2.4}$ days after the discovery of the X-ray outburst on 2021-12-20. These constraints now enable us to explore intrinsic properties of the outflows and accretion in \target.

It is worth noting that, as shown in Figure~\ref{fig:motion} - {\it right}, the model fails to explain the positional scatter of the lobes as observed by MeerKAT between $\sim$ MJD 59300 and 59550. This scatter cannot be described by any simple model that does not account for effects such as energy injection or systematic effects such as uncertain localization of the marginally-resolved lobes due to the presence of hot spots. Thus, it is likely that the statistical uncertainty in the launch date inferred from our model is underestimated. It is also difficult to compare our inferred jet launch date with the date of any contemporaneous X-ray state transition, as \target\ spent a large fraction of its outburst in sun-constraint for many X-ray observatories.

\begin{longtable}[c]{llcc} 
\caption{Prior assumptions and posterior estimates for model parameters. All priors were assumed uniform ($\mathcal{U}$) with limits set to physically meaningful boundaries. The choice of prior for $t_0$ is to encompass the interval from before the detection of the X-ray outburst to after the first detection of the two lobes. Reported posterior estimates are median, 0.16 and 0.84 quantiles. $\widehat{R}$ is the rank-normalized convergence diagnostic \citep{Gelman1992, Vehtari2019} to allow assessment of chain convergence.}
\label{tab:parameters}
\endfirsthead                                          \hline
Parameter                               & Prior                      & Posterior estimates      & $\widehat{R}$\\
\hline
\hline
$\mu_{0,N}$ (mas~d$^{-1}$)               & $\mathcal{U}(0,1000)$     & $33.1\pm1.4$            & 1.00004\\
$\dot{\mu}_N$ ($10^{-2}$ mas~d$^{-2}$)   & $\mathcal{U}(-1000,0)$    & $-3.0\pm0.8$            & 1.00006\\
$\mu_{0,S}$ (mas~d$^{-1}$)               & $\mathcal{U}(-1000,0)$    & $-48.1\pm2.1$           & 1.00006\\
$\dot{\mu}_S$ ($10^{-2}$ mas~d$^{-2}$)   & $\mathcal{U}(0,1000)$     & $11.6_{-1.2}^{+1.3}$    & 1.00007\\
$\theta_\textrm{scatter}$ ($^\circ$)     & $\mathcal{U}(0,90)$       & $6.5_{-0.5}^{+0.6}$     & 1.00003\\
$t_0$ (MJD)                              & $\mathcal{U}(59100,59300)$& $59246.4_{-2.5}^{+2.4}$ & 1.00002\\
\hline
\end{longtable}
\vspace{1cm}

\section{Discussion}\label{sec:disc}
\subsection{Distance to \target\ and its association with \gc}\label{sec:distance}
In order to interpret the proper motions estimated above, we need to constrain the distance to \target. First, we can obtain an upper limit on the distance using our estimates of the initial proper motions of the lobes ($\mu_{0,N}$, $\mu_{0,S}$), using the constraint that they cannot intrinsically be moving faster than the speed of light \citep[e.g.,][]{Fender1999, Fender2006}. Using the posterior samples discussed in \S\ref{sec:motion}, we obtain an upper limit of $d_\textrm{max}=4.3\pm0.2$ kpc, which rules out most of the Galactic bulge.

Second, while \gc\ is within the Galactic plane \citep[located at $\sim$\distance,][]{Hare2018}, with most of the Galactic stellar mass in this direction behind it, the fractional density of X-ray binaries in globular clusters per unit stellar mass is $\sim150$ times higher than in the rest of the Galaxy \citep[e.g.,][]{Clark1975, Heinke2003c}.

Thus, while the probability of a chance coincidence -- and that \target\ could be behind \gc\ -- might have been worth considering in the absence of any other independent constraints, the presence of a tight maximum distance constraint that rejects distances $\gsim 5$~kpc makes this probability negligible. 

In addition, we should then consider whether \target\ could be a foreground object, in front of \gc. While this possibility cannot be ruled out, it is still an unlikely scenario, as there is comparatively little stellar mass in this direction in front of \gc. 

Therefore, we conclude that \target\ is likely located within \gc, at a distance of \distancewer\ as estimated by \citet{Hare2018}. \citet{Hare2018} estimate the distance to \gc\ based on identification of red clump stars in infrared observations. Given the high levels of extinction and crowding, they perform multiple iterations on selection and provide an estimate of 3.3--3.5~kpc, with a more conservative interval of 3.0--3.7~kpc. Thus, based on their work, here we conservatively assume a distance of \distancewer.

\subsection{A relativistic outflow from a globular cluster XRB and implications on the accretion properties}\label{sec:beta}
\begin{figure}
\centering
\includegraphics[width=3.5in]{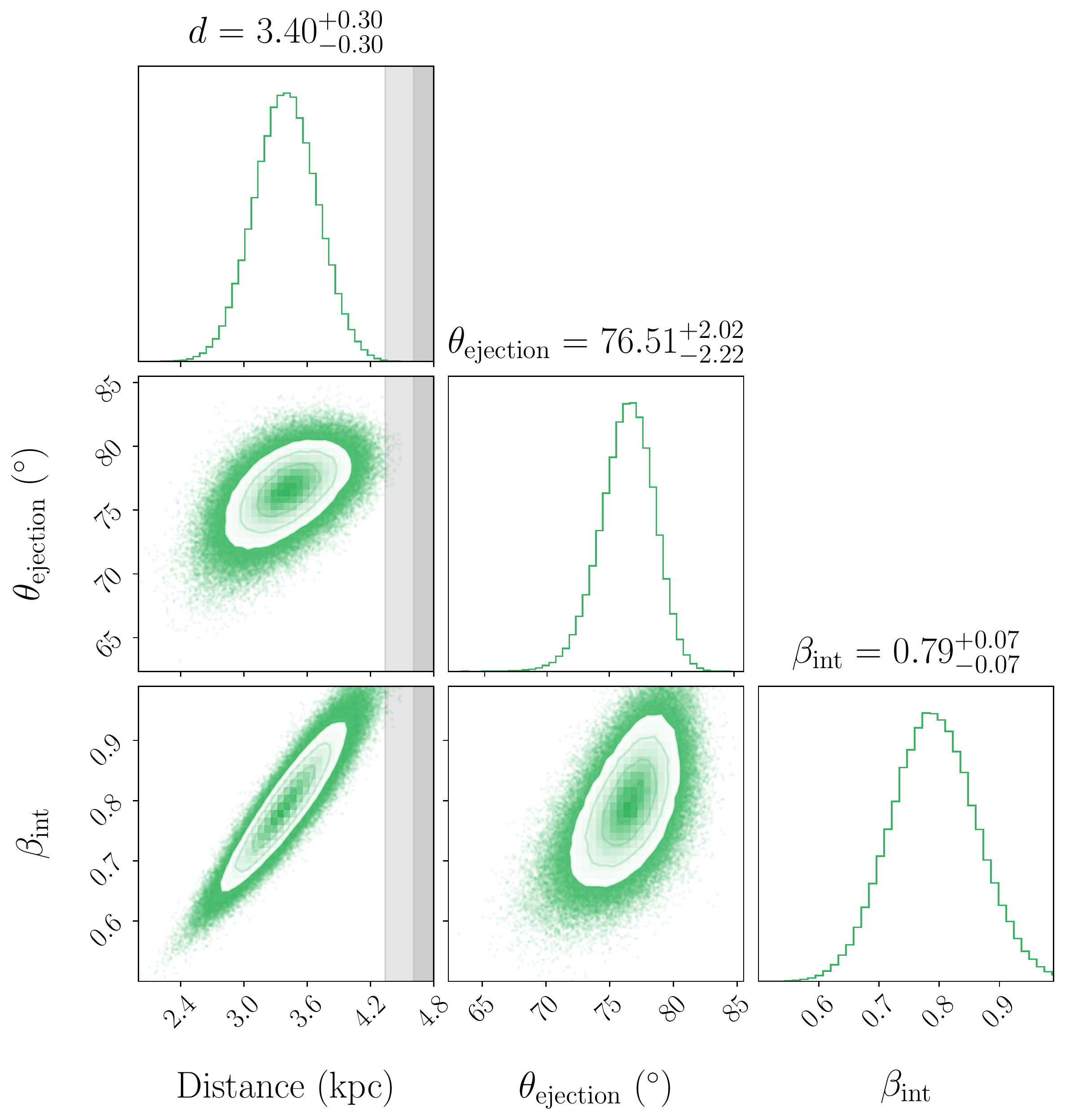}
\caption{Pairwise plots for the ejection angle and the intrinsic speed of ejection ($\beta_\textrm{int}=v/c$, where $c$ is speed of light), estimated using the posterior samples for $\mu_{0,N}$ and $\mu_{0,S}$, and assuming a distance of \distancewer\ \citep{Hare2018}. The shaded light and dark gray intervals in the pairwise panels containing distance indicate the median and 95\% quantile, respectively, on the maximum possible distance as inferred in \S\ref{sec:distance}.}
\label{fig:beta}
\end{figure}

Using the constraints on initial proper motions of the lobes ($\mu_{0,N}$, $\mu_{0,S}$), and assuming that the outflows are intrinsically symmetric, we can estimate the angle of ejection to the line of sight ($\theta_\textrm{ejection}$) and the intrinsic velocity of the lobes ($\beta_\textrm{int}=v/c$) following \citet{Rees1966, Blandford1977, Mirabel1994, Fender2006}. For this purpose, we use our posterior samples for $\mu_{0,N}$ and $\mu_{0,S}$ and estimate $\beta_\textrm{int} \cos \theta_\textrm{ejection}=0.19\pm0.02$. Following that, with a randomly generated sample of distance values assuming a normal probability density function for distance with a mean of 3.4~kpc and a standard deviation of 0.3~kpc (see \S\ref{sec:distance}), we estimate $\theta_\textrm{ejection}=77^\circ\pm2$ and subsequently $\beta_\textrm{int}=0.79\pm0.07$ (Figure~\ref{fig:beta}). 

If we assume there is no significant misalignment between the jet and accretion disk \citep[but see][]{Maccarone2002}, the inferred $\theta_\textrm{ejection}$ indicates a high inclination angle. A high inclination angle would be in tension with the inclination angle of the inner accretion disk estimated by \citet{Pike2022}, who used X-ray spectral modeling to infer a low inclination angle of $\approx 26.4\pm0.5^\circ$ for the inner accretion disk. This tension may indicate that there is a high degree of disk-jet misalignment in \target. However, we note that a high inclination angle would help explain the apparently low X-ray luminosity of the outburst states observed in this system \citep[as highlighted by][]{Pike2022} through obscuration. 

Additionally, \citet{Pike2022} indicate that while their modeling is not particularly sensitive to the inclination of the outer disk, when left to vary independently, it yields a suggestive value of $79_{-11}^{+6}$ degrees. This may suggest that the accretion disk could be warped. If this is indeed the case, the potential alignment between the jet and the outer accretion disk (while the inner disk is misaligned with both) may be indicative of the influence of the outer disk on outflow launching in XRBs \citep[e.g.,][]{Liska2019}. 

Our inferred values of the initial proper motions ($\approx$ 34 and 50 mas~day$^{-1}$) and intrinsic jet speed $\beta_\textrm{int}=0.79\pm0.07$ indicate a relatively slow moving outflow for \target\ as compared to some other XRBs. For example, outflows from GRS~1915+105 \citep{Mirabel1994, Fender1999}, GRO~J1655$-$40 \citep{Hjellming1995}, XTE~J1550$-$564 \citep{Corbel2002}, GX~339$-$4 \citep{Gallo2004}, XTE~J1550$-$564 \citep{Hannikainen2009}, MAXI~J1535$-$571 \citep{Russell2019}, MAXI~J1820+070 \citep{Bright2020} and MAXI~J1348$-$630 \citep{Carotenuto2021} have been observed to be faster, and relatively close to the speed of light\footnote{It is worth noting that distance to many of these systems is not well constrained and very few have measured parallaxes. This can influence estimated intrinsic velocities significantly \citep{Fender2003}.}. In contrast, systems such as SS~433 \citep{Fabian1979}, XTE~J1752$-$223 \citep{Yang2010, MillerJones2011}, XTE~J1908+094 \citep{Rushton2017}, V404~Cygni \citep{MillerJones2019}, EXO~1846$-$031 \citep{Williams2022} have shown slower-moving outflows, with speeds comparable to \target, as estimated in this work. In addition, among the large-scale outflows from XRBs, those observed in \target\ have propagated a relatively small distance at a slow apparent speed - e.g., in contrast with MAXI~J1820+070, MAXI~J1348$-$630, XTE~J1550$-$564, MAXI~J1535$-$571. However, some XRBs have been observed to decelerate on smaller angular scales \citep[e.g., XTE~J1752$-$223, EXO~1846$-$031,][]{MillerJones2011, Williams2022}.

Lastly, it is worth noting that \citet{Pike2022} detected evolving narrow Fe emission lines in their observations of \target\ in December 2020 -- considerably earlier than our first detection of the outflows at the end of February 2021, or the projected outflow launch inferred from our model, at the beginning of February 2021. They discuss jet launching as one possible scenario for the origin of these lines. While we might have found the jets predicted by \citet{Pike2022}, the timeline of events suggests that there could potentially have been multiple jet launching events. However, we find no evidence for multiple sets of jet knots in either the VLA or the MeerKAT observations. In particular, we imaged the VLA observation out to 4.5$'$ (the full-width half-maximum of the VLA primary beam at 10 GHz) and searched along the jet propagation axis for evidence of ejecta that may have been launched earlier than the lobes detected in the VLA observation (Figure~\ref{fig:vlaimg} - \textit{left}) and we found no evidence for the presence of such ejecta.

\subsection{Nature of the system and the compact object}
The likely presence of \target\ in a globular cluster indicates that it is a low-mass X-ray binary (as opposed to a high-mass X-ray binary). Based on its X-ray temporal and spectral properties, particularly the broad Fe line, \citet{Pike2022} speculate that the compact object in this system may be a black hole - while determining the nature of the compact object in XRBs using such features can be uncertain.

Our observations of strong outflows from \target\ also hint at a black hole nature, as similarly powerful (i.e., bright and moving at relativistic speeds) outflows are more commonly seen from accreting black holes than from neutron star systems. However, we can not conclusively rule out a neutron star nature, as accreting neutron stars with powerful outflows have also been observed -- albeit, much less frequently than black holes. For example, such outflows have been seen from the neutron star XRBs Circinus~X-1 \citep{Fender2004b, Tudose2008} and Scorpius~X-1 \citep{Fomalont2001a, Fomalont2001b}, although both these sources differ from \target\ in some of their key characteristics. Circinus X-1 is a very young \citep[$\sim5$\,kyr;][]{Heinz2013} neutron star high-mass X-ray binary system, which are not found in old star clusters. Scorpius X-1 is a neutron star low-mass X-ray binary persistently accreting at close to the Eddington luminosity (in contrast to the transient nature of \target). While it has shown bright relativistic outflows, at their brightest these were $\sim$an order of magnitude fainter than the outflows observed in \target\ at their brightest, despite the distances being comparable.

We detect faint radio emission from a putative compact core (\S\ref{sec:localization}). Considering the disk-jet coupling in accreting systems \citep{Fender2003, lrlx_repo}, the VLA detection and the Swift/XRT observation performed on the same day as the VLA observation would place \target\ below both the black hole tracks and the vast majority of neutron star systems (L$_\textrm{R}\approx3\times10^{26}$~\ergs, and an apparent L$_\textrm{X}\approx10^{35}$~\ergs). However, the X-ray spectrum of that observation and the one performed a few days earlier and reported by \citet{Kennea2021} both indicate a rather soft spectrum (power-law photon index $\geq2$) which suggest that the source was not in a hard state at the time, and thus it would be inappropriate to place it on the radio/X-ray plane. The high inclination angle also can lead to underestimation of the X-ray luminosity, making the source even more underluminous in the radio \footnote{For an example of how inclination angle affects the disk-jet coupling diagnostics, see the discussion on GRS~1747$-$312 in \citet{panurach2021}.}

Thus, while our observations of the outflows from \target\ indicate a behavior consistent with previously observed black hole candidate Low-mass X-ray binaries, the nature of the compact object in this system cannot be definitively confirmed.

It is worth noting that if \target\ does host a black hole, it would be among the very small sample of black holes identified in globular clusters. While globular clusters are expected to produce a large population of black holes through stellar evolution, almost all of these black holes are expected to escape the cluster over a rather short period of time \citep{Sigurdsson1993}. However, the presence of black holes in globular clusters has recently been confirmed through detection of dynamically confirmed ``detached'' (non-accreting) black holes in NGC~3201 \citep{Giesers2018, Giesers2019}. 

To date, there have been no dynamically confirmed black hole XRBs identified in any globular cluster, although over the past two decades, a handful of candidates have been identified in Galactic and extragalactic globular clusters  \citep[e.g., see ][]{Maccarone2007, Strader2012, Chomiuk2013, MillerJones2015, Shishkovsky2018, Dage2019, Dage2020}. All the Galactic candidates among these have been identified in quiescent/faint states (L$_\textrm{X}\leq 10^{34}$~\ergs) and none have so far shown bright (L$_\textrm{X}\geq 10^{35}$~\ergs) outbursts. In contrast, almost all the currently known bright persistent and transient Galactic globular cluster XRBs have been confirmed to contain neutron stars \citep[see][table 7]{Bahramian2022}. Thus, \target\ may be the first outbursting black hole X-ray binary identified in a Galactic globular cluster.

\section{Conclusion}
In this work we presented results of our radio monitoring of the outburst of \target and report the first detection of expanding and decelerating outflows from an XRB in a globular cluster. Using MeerKAT observatory, we monitored large-scale outflows from \target\ regularly for over 500 days, obtaining an extraordinary coverage of such outflows. Using a kinematic model, we constrain initial proper motion of the outflows and constrain possible launch date. We find $\beta_\textrm{int} \cos \theta_\textrm{ejection}=0.19\pm0.02$. Assuming \target\ is located in the globular cluster \gc, at \distance, we determine the intrinsic jet speed, $\beta_\textrm{int}=0.79\pm0.07$, and the inclination angle to the line of sight, $\theta_\textrm{ejection}=76^\circ\pm2^{\circ}$. Additionally, the estimated proper motions in our modeling imply a maximum possible distance of $4.3\pm0.2$ kpc, and indicate comparatively slow moving outflows in contrast with XRBs that have been observed with outflows. Our findings also provide additional circumstantial evidence indicating \target\ could be a black hole X-ray binary, which if confirmed would make it the first such system in a globular cluster to show a transient outburst.

\begin{acknowledgments}
AB thanks Eric W. Koch for helpful discussions. AJT acknowledges support for this work was provided by NASA through the NASA Hubble Fellowship grant \#HST--HF2--51494.001 awarded by the Space Telescope Science Institute, which is operated by the Association of Universities for Research in Astronomy, Inc., for NASA, under contract NAS5--26555. FC acknowledges support from the Royal Society through the Newton International Fellowship programme (NIF/R1/211296). JS acknowledges support from NASA grant 80NSSC21K0628 and the Packard Foundation.

The MeerKAT telescope is operated by the South African Radio Astronomy Observatory, which is a facility of the National Research Foundation, an agency of the Department of Science and Innovation. We acknowledge the use of the Inter-University Institute for Data Intensive Astronomy (IDIA) data intensive research cloud for data processing. IDIA is a South African university partnership involving the University of Cape Town, the University of Pretoria and the University of the Western Cape.

We thank the VLA director and staff for accommodating our DDT request. The National Radio Astronomy Observatory is a facility of the National Science Foundation operated under cooperative agreement by Associated Universities, Inc. We acknowledge extensive use of NASA's Astrophysics Data System Bibliographic Services, Arxiv, and SIMBAD \citep{Wenger2000}.
\end{acknowledgments}

%

\vspace{5mm}
\facilities{VLA, MeerKAT} 


\software{Astropy\citep{AstropyCollaboration2013,AstropyCollaboration2018},
       CASA \citep{CASA2022},
       CMasher \citep{cmasher2020},
       Corner \citep{ForemanMackey2016},
       IPython \citep{Perez2007}, 
       Jupyter \citep{Kluyver2016}, 
       Matplotlib \citep{Hunter2007}, 
       Numpy \citep{oliphant2006,vanderWalt2011},
       PyMC \citep{Salvatier2016},
       SAOImage DS9 \citep{Joye2003},
       Scipy \citep{Virtanen2020}.}





\bibliography{all_references}{}
\bibliographystyle{aasjournal}



\end{document}